\documentclass[journal]{IEEEtran}
\usepackage{mathptmx}
\usepackage{mathtools}
\usepackage{hyperref}
\usepackage{cleveref}
\usepackage{graphicx}
\usepackage{multirow}
\usepackage{longtable}
\usepackage{booktabs} 
\usepackage{color}
\usepackage{siunitx}

\usepackage{latexsym}

\DeclareMathOperator{\arcsinh}{arcSinh}

\usepackage[normalem]{ulem}
\usepackage{xcolor}
\usepackage{bm}
\usepackage{commath}
\usepackage{amsmath}
\usepackage{caption}
\usepackage{subcaption}
\usepackage{csquotes}
\usepackage{gensymb}
\usepackage{multirow}
\usepackage{array}
\usepackage{adjustbox}
\usepackage{booktabs}
\usepackage{soul,xcolor}

\newcommand{\ans}[1]{\color{black}#1 \color{black}}

\begin{document}

\title{An Optical Parametric Amplifier via $ \chi^{(2)} $ in AlGaAs Waveguides}

\author{Zhizhong~Yan, Haoyu~He, Han~Liu, Meng~Iu, Osman~Ahmed, Eric~ Chen, Youichi~Akasaka, Tadashi~Ikeuchi and~Amr S. Helmy, ~\IEEEmembership{Fellow,  ~OSA}
\thanks{Zhizhong Yan, Han Liu, Meng Iu, Osman Ahmed, Eric Chen and Amr S. Helmy are
with the Edward S. Rogers Sr. Department of Electrical and Computer Engineering, University of Toronto,  10 King’s College Road, Toronto, Ontario, Canada M5S 3G4. \newline e-mail: a.helmy@utoronto.ca.}
\thanks{Youichi Akasaka and Tadashi Ikeuchi are with Fujitsu Laboratories of America, 2801 Telecom Parkway, Richardson TX 75082 USA.}
\thanks{Manuscript received February, 2022.}}
\markboth{IEEE Journal of Lightwave Technologies,~Vol.~x, No.~x, February~2022} %
{Yan \MakeLowercase{\textit{et al.}}: An Optical Parametric Amplifier via $ \chi^{(2)} $ in AlGaAs Waveguides}

\maketitle

\begin{abstract}
We report parametric gain by utilizing $ \chi^{(2)} $ non-linearities in a semiconductor Bragg Reflection Waveguide (BRW) waveguide chip. Under the two-mode degenerate type II phase matching, it can be shown that more than 18 dBs of parametric gain for both TE and TM modes is tenable in 100s of micrometers of device length. Polarization insensitive parametric gain can be attained within the 1550 nm region of the spectrum. These AlGaAs BRW waveguides exhibit sub-photon per pulse sensitivity. This is in sharp contrast to other types of parametric gain devices which utilize $ \chi^{(3)} $, where the pump wavelength is in the vicinity of the signal wavelength. This sensitivity, which reached 0.1~photon/pulse, can usher a new era for on-chip quantum information processing using compact, micrometer-scale devices.
\end{abstract}

\begin{IEEEkeywords}
Nonlinear Optics, Ultra-fast optics, Quantum Optics, Phase Insensitive amplifier (PIA), Quantum Noise,
Optical parametric amplifier, Polarization.
\end{IEEEkeywords}

\IEEEpeerreviewmaketitle

\section{Introduction}
\IEEEPARstart{T}{he} first parametric amplifier was successfully implemented in 1965 \cite{Wang1965}. Since then, parametric devices have played a pivotal role in quantum optical devices, optical signal processing and optical communications systems. In 1993, a table-top optical parametric amplifier (OPA) using bulk KTP nonlinear crystal~\cite{Levenson1993} was first demonstrated. OPAs with gain factors as high as 70~dB~\cite{Hansryd2002} ($10^{-4} \rm{dB \cdot mm^{-1}}$, \ans{CW pump}) can be obtained and are predominantly implemented in optical fibers using $\chi^{(3)}$ non-linearities owing to the low fiber losses. There also exist several approaches to realize on-chip integrated OPAs using $\chi^{(3)}$.

Silicon waveguides which utilize four-wave mixing (FWM) have demonstrated parametric gain in the telecommunication band \cite{Foster2006, Salem2008}, and also at longer wavelengths \cite{Liu2010, Kuyken2011, Liu2012}. Chalcogenide waveguides \cite{Schroder2013} have also been used to demonstrate OPAs with peak gain values up to 30 dB ($3 \rm{dB.mm^{-1}}$)\cite{Lamont2008}. Compound semiconductors can be used as stand alone amplifiers or can be integrated on Si. They offer advantages when used to implement OPAs. Using FWM  to obtain parametric amplification in GaAs quantum wells was first reported in 1992 by Kim et al.~\cite{Kim1992}. Recently, AlGaAs-based waveguides have been utilized to achieve parametric gain through FWM \cite{Dolgaleva2015, Wathen2014}. Frequency combs have been recently implemented in AlGaAs on insulator using highly resonant ring cavities\cite{Chang2020}.  The limiting factor in these is primarily due to the pump losses induced by two-photon and free carrier absorption. The highest FWM gain of 42.5 dB in the telecommunication band has been obtained from an ultra-silicon-rich nitride waveguide (USRN) with a length of seven millimeter \cite{Ooi2017}, which represents one of the highest reported gain values per unit length (~$6 \rm{dB \cdot mm^{-1}}$, \ans{CW pump}).

Second-order non-linearities $\chi^{(2)}$ offer various advantages for parametric devices, including a larger optical non-linearity when compared to $\chi^{(3)}$ and a large frequency spacing (over $\SI{100}{THz}$) between the signal/idler and the pump. The large frequency difference facilitates the efficient separation of the strong pump from the weaker signal/idler waves. In addition, the large frequency spacing also reduces pump-induced noise into the signal/idler spectral region. Bulk optics-based (KTP crystal) $\chi^{(2)}$ OPAs can provide nearly 9 dB of gain (~1$\rm{dB \cdot mm^{-1}}$, \ans{CW pump}) \cite{Levenson1993}.

Periodically poled lithium niobate (PPLN) is the most commonly used material to construct $\chi^{(2)}$ based OPAs. PPLN waveguides are usually several centimeters in length. Recently, Zn-doped $ \rm{LiNbO}_3 $ core layer PPLN waveguides have been introduced \cite{Umeki2011, Umeki2013a, Ishimoto2016}, where they were shown to tolerate high pump power levels exceeding $ 10^3 $~mW (1~$\rm{dB \cdot mm^{-1}}$ \ans{CW pump}). The parametric gain obtainable using such PPLN waveguides is typically around 10 to $\SI{20}{dB}$~\cite{Ishimoto2016}. \ans{ Also recently substantial parametric  gain per unit length that has been demonstrated in nanophotonic lithium niobate waveguides \cite{Ledezma:22}, dispersion engineered PPLN nanowaveguides \cite{Jankowski:22}}.

Thus far, it has been challenging to achieve appreciable gain above 1~dB using continuous-wave pumping for AlGaAs waveguides~\cite{Wathen2014, Ravaro2007, Ozanam2014}.  In this work, we demonstrate the first parametric amplification that has record gain $>>$ 18~$\rm{dB \cdot mm^{-1}}$ using a femtosecond pulsed pump laser and type-II phase-matching configuration. The main advantage of a type-II phase-matching configuration is the ability to achieve near polarization-insensitive gain over a part of the gain spectrum. \ans{As a comparison, we have summarized previous OPAs that are implemented in integrated platforms in Table \ref{S-Tab:OPASummary}}. 
\section{Parametric Gain}
\begin{figure}[t]
	\centering
	\includegraphics[width=1 \columnwidth]{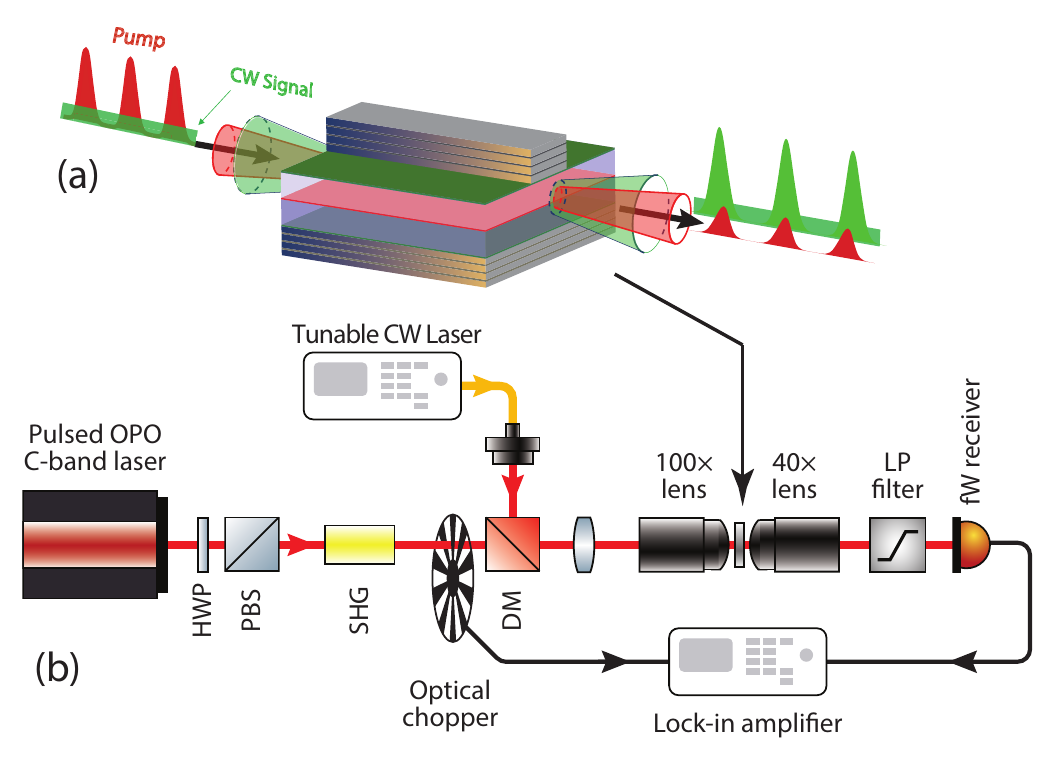}
	\caption{\textbf{The experimental setup and device structure.} (a) Operating principle of the the device which is based on a Bragg waveguide. The pulsed pump (red) and CW signal (green) co-propagate after being focused into the waveguide core by a shared microscope objective. The amplitude of pump is strongly attenuated at the output facet of the waveguide. The amplified signal is in the pulsed form that is superimposed on the CW background.
		(b) Experimental setup for measuring the gain from the AlGaAs waveguide. A C-band OPO is tuned to pump a second-harmonic generation (SHG) nonlinear crystal (BiBO) for generating the pump. The OPO pump power can be adjusted by a HWP and PBS combination. The pump wavelength is centered at 777 nm with  $\SI{120}{\femto\second}$ pulse width. The main optical components include FC: fiber collimator; HWP: half wave plate; PBS: polarizing beam splitter; DM: dichroic mirror; LP: long-pass filter; fW receiver is a Newport femtoWatt detector. An optical chopper is placed in the pump path, providing the reference frequency for a lock-in amplifier (model SR830).}
	\label{Fig-1}
\end{figure}
The structure of the device, where we implemented the OPAs is illustrated in Fig.~\ref{Fig-1}~(a), the BRW AlGaAs waveguide has a length of 1 mm, an etch depth of 3.6 $\mu$m and a ridge width of 2 $\mu$m. \ans{The device detailed design including the Bragg layer thickness, as well as the modal profiles had been reported previously in \cite{Abolghasem2012}.}

The pump field is attenuated as it propagates along the waveguide length due to several factors associated with optical losses. The total optical loss ($ \alpha _t $) arises mainly due to linear ($ \alpha _L $) and nonlinear ($ \alpha _{NL} $) losses such that, 
\begin{equation}\label{eq:alpha_t}
	{\alpha _t} = {\alpha _{NL}} + {\alpha _L}
\end{equation}
The linear loss  is independent of the field intensity, but highly dependent on wavelength. The propagation losses at the range of signal wavelength were measured to be $\sim$4 cm$^{-1}$. 

The nonlinear loss typically takes place in semiconductor materials due to the dissipative part of the nonlinear coefficient. The main nonlinear loss of the pump used in our samples is due to two-photon absorption (TPA). The TPA effect adds extra loss through increasing the imaginary part but has a minor impact on the real part of refractive index. This was confirmed through measurement: the phase matching wavelength was monitored as a function of pump power in a  second harmonic generation experiment. No pump power dependence was observed indicating that for the ramge of pump powers used the effective index change due to self phase mudation is not significant.

Carrier induced losses are also unlikely to play a dominant role in our devices: Unlike  active semiconductor devices, where externally injected carriers, through current injection, can significantly change the refractive index of the waveguide guiding area, the free carrier absorption in the guiding area of our devices, which are generated due to a pump with a low duty cycle, has negligible effect on the loss of these devices. This is because in our experiments, the carrier generation is induced by the short pulse of the pump, which has a  duty cycle of $ 10^{-5}$.

The parametric gain is obtained using a type-II parametric process in a our devices \cite{Han2009, Han2010a}\ans{. The structure from which these devices where fabricated exhibits} very high $ \chi^{(2)} $ non-linearity ($ 10^4 \% \cdot \rm{W}^{-1} \cdot \rm{cm^{-2}} $) as witnessed in previous second harmonic difference frequency and sum frequency experiment \cite{Abolghasem2009}. The pump and signal input to the parametric gain process are generated separately and the amplification is phase-insensitive \cite{Caves1981}. For type-II phase matching, the device inputs can be any combination of TE or TM signal power levels with arbitrary relative phases; whereas, for type I phase matching, only a TE input is permissible. Our type-II phase-matched OPA can be operated across the C and L wavelength bands.

The experimental setup displayed in Fig.~\ref{Fig-1}~(b) includes a strong short pulse pump, set in a transverse electric (TE) mode for the type-II phase-matching condition. The femtosecond pump is produced by an optical parametric oscillator (OPO), whose pulse duration is less than $\SI{120}{\femto\second}$ (80 MHz repetition rate) with a center frequency $\omega_p$. The OPO is frequency doubled in a 1~mm thick BiBO crystal to provide the pump, which had a wavelength around 780nm with several THz bandwidth. A CW tunable laser in the C- and L-bands of wavelength is used to produce the narrow band signal. The signal strength is varied by a variable optical attenuator (VOA). 

The pump coupling is maximized for all measurements, while the signal path is optimized to be collinear with the pump path, as shown in Fig.~\ref{Fig-1}. Both use the $100\times$ microscope objective for coupling in the amplifiers. The pump is strongly attenuated by two-photon absorption (TPA) in the amplifier waveguide providing an effective length shorter than the physical chip length, where the nonlinear interaction takes place. We have included detailed calculations on how the nonlinear loss impacts the total optical loss and effective length (Eq.~(\ref{eq:alpha_t})) in Appendix A. We found that the effective length is about 400~$ \mu $m at 1 mW pump average power; and reduces to near 50~$ \mu $m at 15 mW pump power, which accounts for near 5\% of the waveguide physical length. The linear loss plays only minor role in determining the effective length.

A critical step to characterize the OPA gain is to accurately determine the \ans{amplifier input power level to the OPA}. However, this is experimentally challenging for several reasons. First, because the input objective lens is only optimized for pump coupling, part of the input signal light will be coupled into higher order spatial mode that experience no amplification. Second, the pump light with duration of approximately 120 fs and 80MHz repetition rate, only amplifies a small fraction of the CW signal which is temporally overlapping. Mathematically this can be understood as the pulsed OPA process consisting of multiple channels of a two mode squeezing operation over different pairs of time frequency(Schmidt) modes in TE and TM polarization. 

\ans{In theory, any given TE and TM polarization signal can be decomposed into a set of parallel Schmidt modes. When this set of parallel modes are amplified by the OPA, the amplified output modes together form the output signal.  One can prove that the a single Schmidt mode for TE and TM polarization experiences the maximal gain. We also studied the OPA under test, and theoretically found that this OPA is operated at near single Schmidt mode (See Appendix C).}

\ans{To determining the value of $ P^{TE}_0,P^{TM}_0 $ for TE/TM modes,} we adopted a quantum optical approach to determine the OPA input power $ P^{TM}_0,P^{TE}_0 $ for both input polarizations which experience the amplification. The detailed experimental steps and analysis can be found  in the Appendix C, subsection C.

\ans{Subsequently, } we  increased the pump power ($ P_p $); and measured the output power for different signal wavelengths $ \lambda_s $. The OPA output power at the OPA length $ L $  inside the waveguide facet is expressed as $ P^{p}_S(L) = P_0^{p} { {{\cosh }^2}r^p}  $, where $P_S^p  (L) $ denotes the power at the output facet in TE and TM modes, respectively. Note that the spontaneous term in the Eq.~(\ref{eq.appendix.PsL}) has been removed during the extraction of the parametric gain of the OPA. The output power is then related to the input power by the polarization-dependent OPA gain: 
\begin{equation}\label{eq:gab}
	{g^p}\left( {{\lambda_s, P_p}} \right) = P^{p}_S(L) / P_0^{p}  \\
\end{equation}
Eq.~(\ref{eq:gab}) suggests how for the TE and TM modes, the gain value can be experimentally determined.
Then the OPA gain $ {g^{TE}}\left( {{\lambda_s, P_p}} \right) $ and $ {g^{TM}}\left( {{\lambda_s, P_p}} \right) $ are defined as functions of the pump power ($ P_p $) and the signal wavelength $ \lambda_s $ in different polarization TE  and TM  modes. In general,  TE and TM gains are not necessarily balanced. The measured OPA gain now becomes a function of both input narrow band signal wavelength $ \lambda_s $, as well as the pump power $ P_p $. 

\begin{figure}[t]
	\centering
	\includegraphics[width=0.99 \columnwidth]{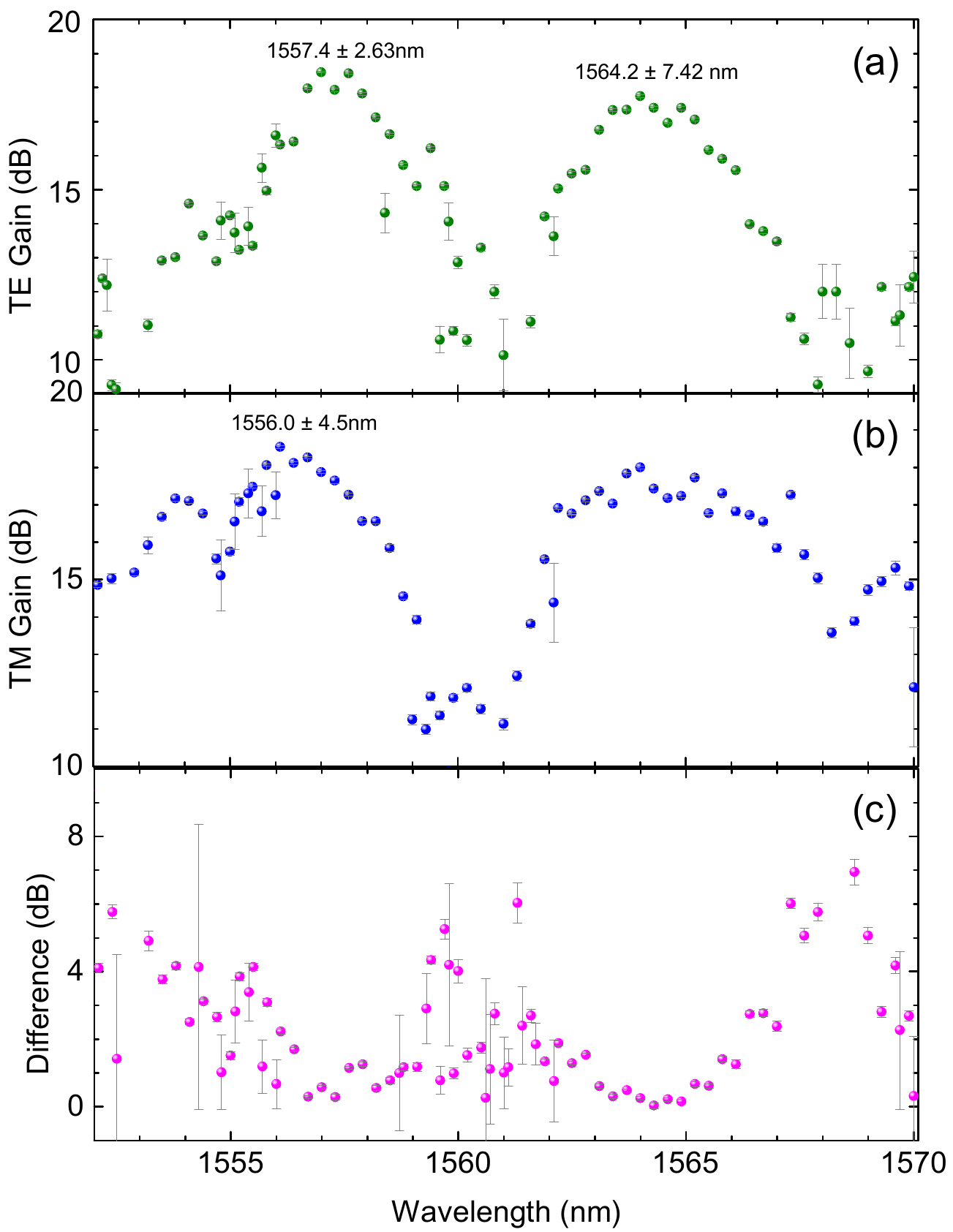}
	\caption{\textbf{The OPA gain measurement results.} The device gain is measured in the waveguide when the signal power $P_S(0)$  is launched in the TE and TM modes, respectively. The signal wavelength is varied from $\SI{1552}{nm}$ to $\SI{1570}{nm}$. The pump is in the TE mode for type-II phase matching. The pump power is set to 15 mW. (a) is the amplified signal experimental result when the input is in the TE mode only. (b) The result with a TM signal input only. (c) The gain difference is measured in a waveguide as the signal wavelength is varied from $\SI{1552.0}{nm}$ to $\SI{1570.0}{nm}$. The degeneracy point of signal and idler is estimated to be located at 1564 nm, given the broadband nature of pulsed pump. }
	\label{fig:Fig-4}
\end{figure}
\begin{figure}[t]
	\centering
	\includegraphics[width=0.9 \columnwidth]{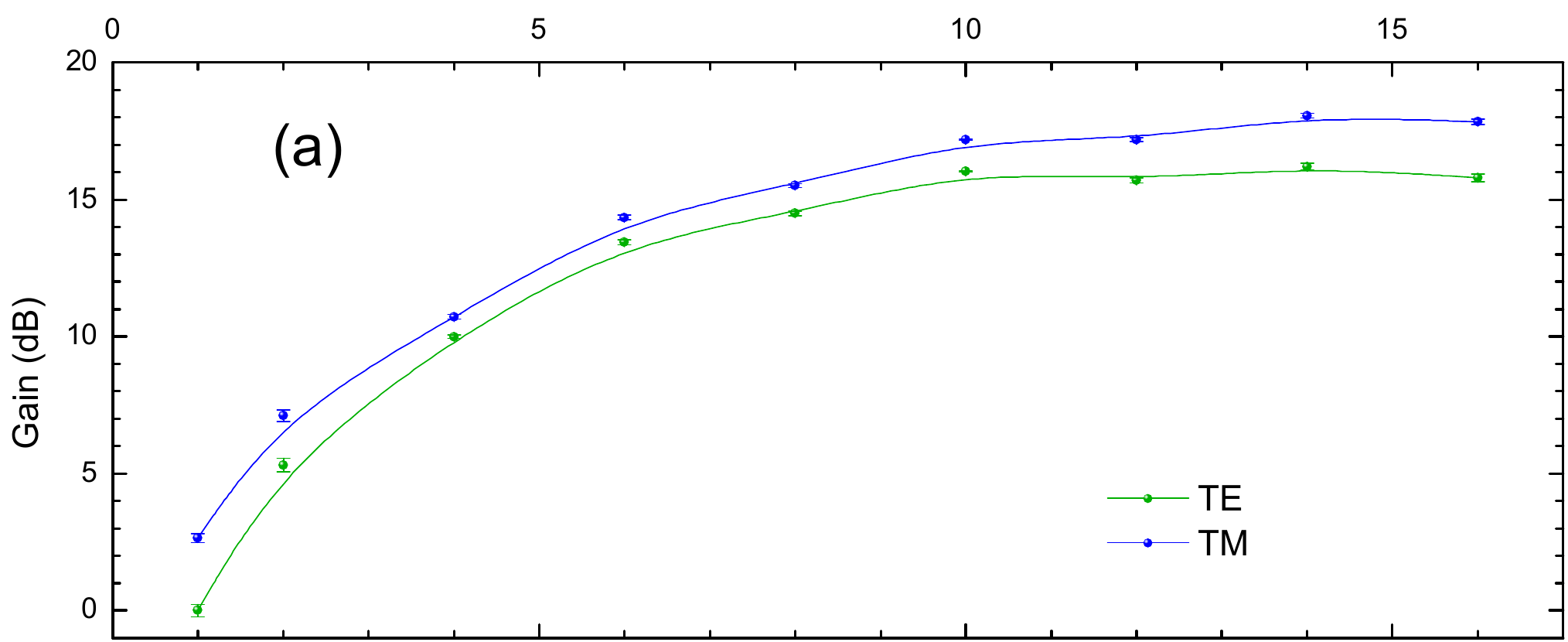}
	\includegraphics[width=0.9 \columnwidth]{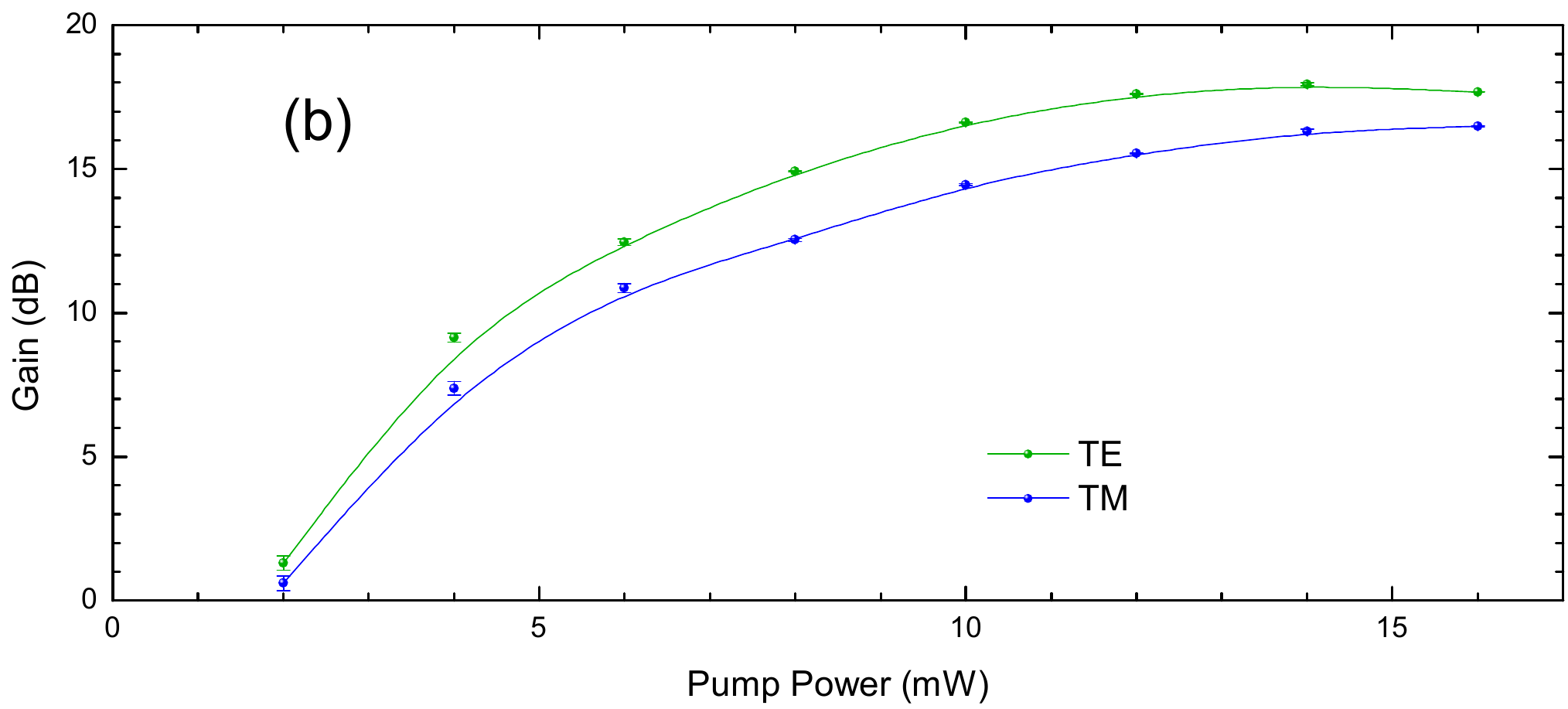}
	\caption{\textbf{The gain polarization insensitivity measurement.} The gain across the 16 mW pump power range, measured for both TE and TM input polarization, while the wavelength was maintained constant. The pump power is scanned from $\SI{0}{mW}$ to $\SI{16}{mW}$. The gain difference in general is appreciable, but at some wavelengths (a) $\lambda_s=\SI{1561.6}{nm}$ and (b) $\lambda_s=\SI{1569.3}{nm}$ the gain difference between the two input states polarization is less than 3 dB.}
	\label{fig:Fig-5}
\end{figure}

Figs.~\ref{fig:Fig-4}~(a) and (b) show  the gain for different input polarization at different frequencies for the TE and TM input polarization states, respectively. We set the pump power to be constant at 15~mW to eliminate any power contribution to the gain imbalance. The difference between the two gain coefficients within the same wavelength range and at the same pump power is displayed in Fig.~\ref{fig:Fig-4}(c). We have identified two regions where there exists near-zero dB gain difference between the TE and TM inputs.

We also plotted the gain difference between the two different input polarizations as a function of pump power. The signal wavelength was scanned from $\SI{1550}{nm}$ to $\SI{1570}{nm}$. We found two points within the wavelength range, namely 1561.6 nm and 1569.3 nm, where the TE-TM gain difference remains  less than $\SI{3}{dB}$ across the entire pump power range tested, which reaches $\SI{16}{mW}$. The results are plotted in Figs.~\ref{fig:Fig-5}~(a) and (b).  \ans{There is also near degeneracy near 1570 nm, however we did not include the 1570 nm region as that wavelength was at the end of the tuning range for our laser and we had no chance to probe the region further. However there would be more degeneracy regimes as we scan wider range. By using coupled Nonlinear Schrödinger's Equations (NLSE) \cite{Han2010a, Han2010}, the parametric gain of the structure can be studied (detailed in Appendix A). The 18dB gain experimental result agrees with the numerical simulation analysis in \cite{Yan:22}. Further increase of the pump power beyond 15mW will negate the gain due to the strong  nonlinear loss.}

\begin{figure}[t]
	\centering
	\includegraphics[width=0.8 \columnwidth]{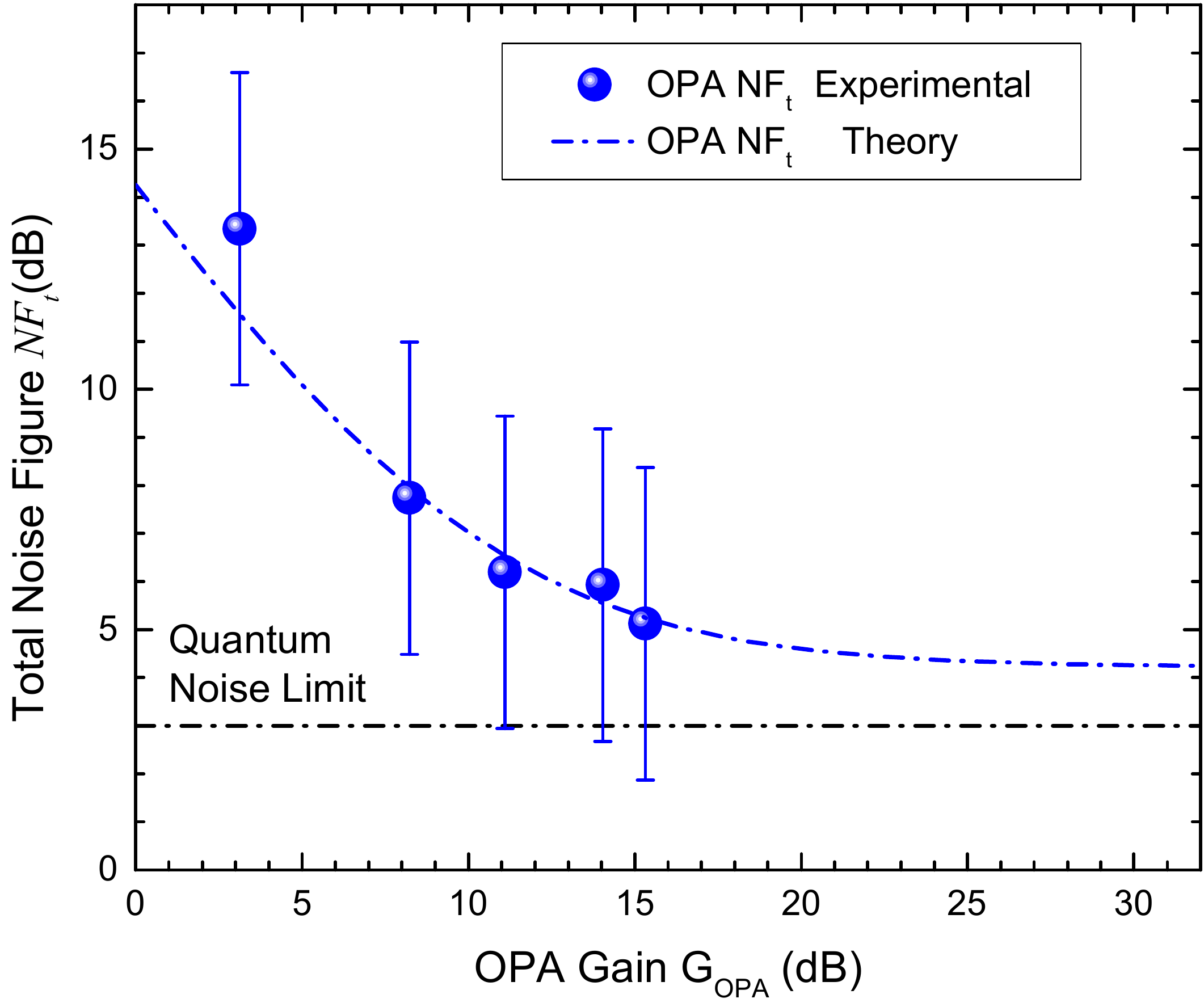}
	\caption{\textbf{The noise figure measurement of the OPA.} The noise figure of the parametric amplifier measured with type-I phase matching, where the pump is TM-polarized. The signal is set in the TE mode only.}
	\label{fig:Fig-6}
\end{figure}

\section{Noise Figure and Sensitivity}
The noise performance of chip-based OPAs has seldom been reported. We have examined the noise performance of these devices by measuring the amplifier noise figure (NF) and thereby examining the photon quantum fluctuations with respect to the average input and output photon numbers. Since the quantum nature of the photon fluctuation can be considered as white noise in nature, this relaxes the requirements for the detector bandwidth to characterize the noise figure in our approach.

For these measurements, type-I phase matching was utilized as it offers a single polarization input and a single polarization output in contrast to type-II configuration, thereby making the measurement feasible. In the type-I configuration, the signal is in the TE polarized mode while the pump is in the TM polarized mode. A lock-in amplifier was used to measure the noise fluctuations. The setup is the same as Fig.~\ref{Fig-1}(b) except that the pump is switched into the TM mode.

The external noise figure $ NF_t $ is measured at the system level and includes all of the optical losses based on the measured $\eta = \SI{4.8}{\percent}$ optical efficiency including all optical and detector losses. The intrinsic noise figure $ NF_{OPA} $ is the noise performance that can be extracted from $ NF_t $ in the absence of all optical losses. Therefore, using Friis' formula~\cite{Levenson1997} 
\begin{equation}\label{eq:NFt}
	NF_t = NF_{OPA} - {1 \over {G_{OPA}}} + {1 \over {\eta {G_{OPA}}}}
\end{equation}	
where $ G_{OPA} $ is the OPA gain. For this measurement we needed to utilize type I phase matching. By measuring $ NF_t $ we are able to find $ NF_{OPA} $ to be as low as 4.7~${\rm{dB}}$ at the maximum tenable gain of 16~$\rm{dB}$. The measured $ NF_{OPA} $ with respect to the gain is plotted in Fig.~\ref{fig:Fig-6}. The dashed line indicates the quantum noise limit (3~dB) of a phase insensitive amplifier.
\begin{figure}[t]
	\centering
	\includegraphics[width=0.9 \columnwidth]{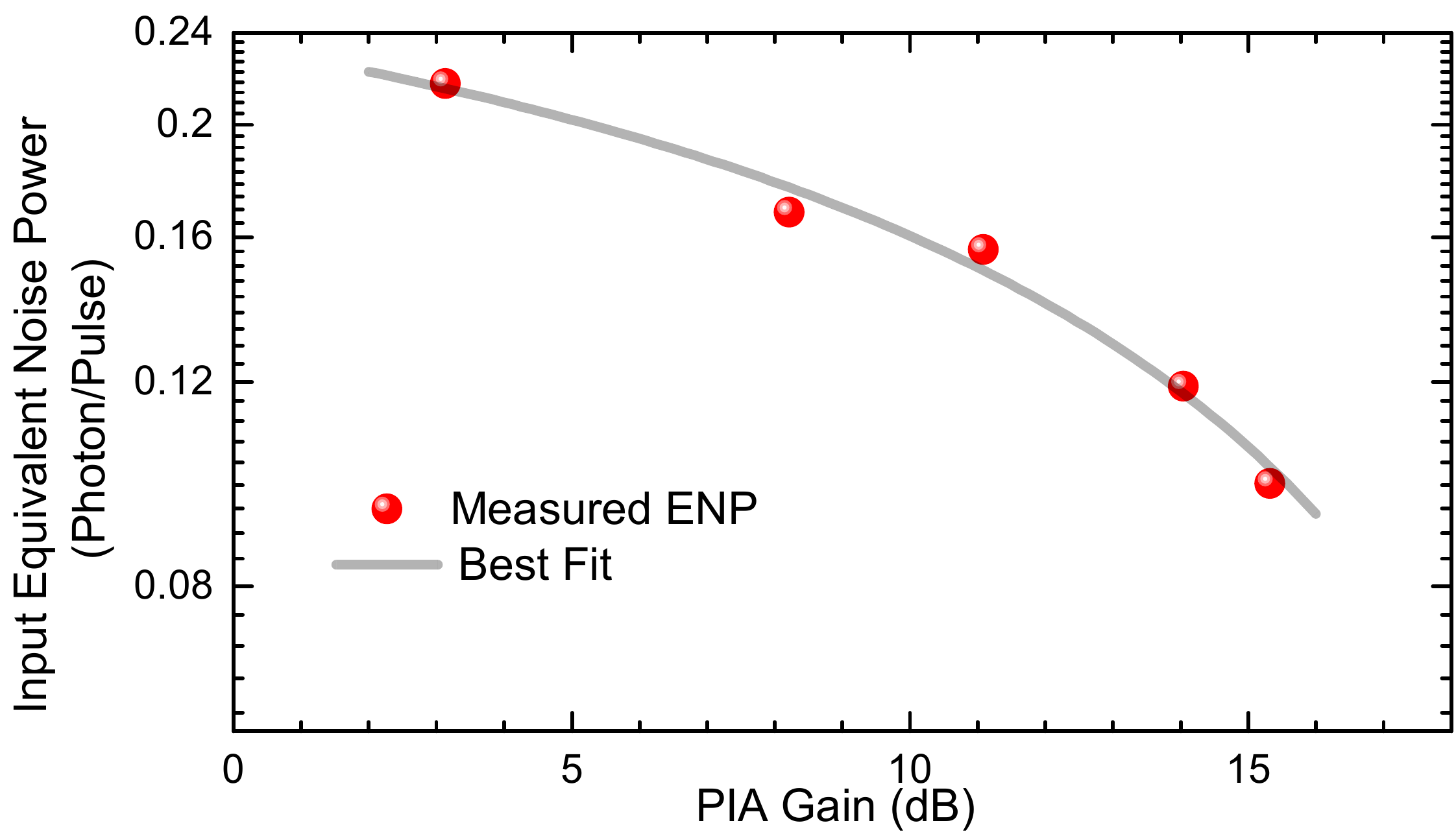}
	\caption{\textbf{The OPA sub-single photon input sensitivity.} The input equivalent-noise-power measured with type-I phase matching, where the pump is TM-polarized. The signal is set in the TE mode only. The ENP is converted into number of photons per pulse via Eq.~(\ref{eq:sensitivity}).}
	\label{fig:Fig-7}
\end{figure}

The OPA sensitivity is another important figure of merit that determines its suitability for numerous domains of applications. 

We quantify the sensitivity as the input noise power, which is also known as the the input equivalent noise power (ENP). The ENP is defined as the minimum signal input \ans{signal} power that is comparable with the noise power \cite{Richards1994} \ans{level at the OPA input.} The measurement of ENP is obtained by directly measuring the spontaneous output power when the input signal is blocked. 

For $\chi^{(2)}$ based parametric devices the pump wavelength is several hundreds of THz away from the signal. The pump has minimal overlap with the signal, which enables more effective filtering and hence enables the single photon sensitivity, which we are able to measure. This is in strong contrast to  $\chi^{(3)}$ based FWM OPAs, where the pump and signal are usually in the same wavelength band. 

Fig.~\ref{fig:Fig-7} shows the experimental results of the input the ENP in terms of the number of photons per pulse, which is indicative to the sensitivity of the amplifier. We have found that the ENP is 0.22 photon per pulse at a gain of 3.1 dB and 0.1 photon per pulse at 15.3 dB gain. The ENP value is dictated by the spontaneous emission power $P_{sp}$ as
\begin{equation}\label{eq:sensitivity}
	ENP = P_{sp} + NF + IL - G_{OPA}.
\end{equation} 
where $ENP$ and $P_{sp}$ are expressed in units of dBm. The $NF$ is the intrinsic noise figure we measured and discussed earlier in this section. The term $IL$ is the optical insertion loss of the device, which has been experimentally determined for our devices to be 13.2 dB. 

From Eq.~(\ref{eq:sensitivity}), we anticipate the ENP to decrease, and consequently the amplifier sensitivity to increase with gain until a certain gain value, where the $P_{sp}$ dominates. Above such gain value, the ENP will reverse direction and be reduced. It is significant to note that the sensitivity of our OPA has been consistently sub-photon per pulse across the entire gain space. 

\section{Conclusion}
The OPA devices reported here utilize two-mode degenerate type II phase matching to show that more than 18 dBs of parametric gain for both TE and TM modes. The devices also offers regions of the spectrum, where the parametric gain can be polarization insensitive. More importantly, the sub-single photon sensitivity can be of great utility to several applications. Most notable this sensitivity can offer appreciable gain to low flux, high purity, low noise input coherent states. We anticipate that OPAs based on $\chi^{(2)}$ will find many applications not only in classical communication, but also in quantum communications\cite{Agarwal2014}. \\

\ans{ We have numerically solved the coupled Nonlinear Schrödinger's Equations (NLSE) to study the parametric gain \cite{Han2010a, Han2010}. The maximal gains measured here agrees with the the numerical result in \cite{Yan:22}, which also showed that under ultra-short, pulsed pump, the nonlinear loss induced by two-photon absorption (TPA) is dominant and responsible for the main optical loss. Further increase of the pump power will be counteracted by this nonlinear loss. Given the device structure and operating wavelength, numerical calculations indicate that parametric gain should be no larger than 30 dB range \cite{Yan:22}. However, we anticipate that the parametric gain can be improved in longer pump wavelength as the TPA coefficient reduces.}

\section*{Acknowledgment}
This work is supported by Natural Sciences and Engineering Research Council of Canada (NSERC).

\renewcommand{\thefigure}{\thesection. \arabic{figure}}
\renewcommand{\theequation}{\thesection. \arabic{equation}}
\setcounter{equation}{0}
\setcounter{figure}{0}
\appendices

\section*{}
\begin{table}[htbp]
	\footnotesize 
	\centering
	\caption{Summary of optical parametric amplifiers using different kinds of nonlinearity on a variety of material platforms.}
	\begin{adjustbox}{angle=270}
		\begin{tabular}{p{18em}|p{7em} |p{4em}|c|p{10em}|p{5em}|c|c|r}
			\hline
			\hline
			Platform & Wavelength (nm) & Nonlinear process & \multicolumn{1}{p{5.0em}|}{Parametric Gain (dB)} & Pump CW/Pulsed & \multicolumn{1}{p{5em}|}{Device Length (mm)} & \multicolumn{1}{p{5.5em}|}{Gain/length (dB/mm)} & \multicolumn{1}{p{5em}}{Reference} \\
			\hline \hline
			Crystalline silicon (c-Si) waveguide & Telecom & $\chi^{(3)}$ & 5     & Pulsed/ 1nm 75 MHz rep rate & 6.4   & 0.8   &  \multicolumn{1}{p{4.em}}{\cite{Foster2006}} \\
			\hline
			Crystalline silicon (c-Si) waveguide & \multicolumn{1}{l|}{2173} & $\chi^{(3)}$ & 50    & Pulsed/ 2ps 76 MHz rep rate & 40    & 1.3   &  \multicolumn{1}{p{4.335em}}{\cite{Salem2008}} \\
			\hline
			CMOS-compatible hydrogenated amorphous (a-Si) silicon waveguide & Telecom & $\chi^{(3)}$ & 26.5  & Pulsed/0.67 nm, 10 MHz & 11    & 2.4   &  \multicolumn{1}{p{4.335em}}{\cite{Liu2010}} \\
			\hline
			CMOS compatible Silicon nitride - USRN:$ \rm{Si_7N_3} $ & Telecom &  $\chi^{(3)}$  & 42.5  & Pulsed /500fs  & 7     & 6     & \multicolumn{1}{p{3em}}{\cite{Ooi2017}} \\
			\hline
			Chalcogenide waveguides PSA & Telecom & $\chi^{(3)}$ & 7     & Pulsed & 65    & 1.08$ \times 10^{-1} $ &  \multicolumn{1}{p{4.335em}}{\cite{Schroder2013}} \\
			\hline
			Chalcogenide waveguides PIA & Telecom & $\chi^{(3)}$ & 30    & Pulsed & 65    & 4.62$ \times 10^{-1} $ &  \multicolumn{1}{p{4.335em}}{\cite{Lamont2008}} \\
			\hline
			Zn-doped LiNbO3 core layer PPLN waveguides & Telecom & $\chi^{(2)}$ & 12.8  & CW    & 50    & 2.56 $ \times 10^{-1} $&  \multicolumn{1}{p{4.335em}}{\cite{Umeki2011}} \\
			\hline
			dispersion-engineered PPLN nano-waveguides & around 2 $\mu$m & $\chi^{(2)}$ & 30  &  Pulsed/75 fs    & 6    &  10 & \multicolumn{1}{p{4.335em}}{\cite{Ledezma:22}} \\
			\hline
			dispersion-engineered PPLN nano-waveguides & 1.7-2.7 $\mu$m & $\chi^{(2)}$ & 88  &  Pulsed/80 fs, 100 MHz    & 6    &  14.6 & \multicolumn{1}{p{4.335em}}{\cite{Jankowski:22}} \\
			\hline
			Our Device (AlGaAs BRW) & Telecom & $\chi^{(2)}$ & 18    & Pulsed/70 fs, 80 MHz & 1     & 18         \\
			\hline \hline
		\end{tabular}%
	\end{adjustbox}
	\label{S-Tab:OPASummary}%
\end{table}%

\section{Estimation of The Total Effective Length $ L^t_{\rm{eff}} $ and Nonlinear Length $ L_{\rm{NL}} $}
The overall optical loss is comprised of two contributions, i.e. linear loss (the propagation loss) and nonlinear loss. The former is not depending on the pump power; whereas the latter is nonlinear and depends on pump intensity. 

The nonlinear loss in our BRW waveguide is primarily due to the carrier-induced two-photon absorption (TPA) effect. TPA is associated with the third order nonlinearities. The response time-scales of TPA is nearly instantaneous \cite{Aitchison1997} (on the order of $ 300 \times 10^{12} $~\si{Hz}, or $ 10^{-15} $~\si{s}). In strong contrast, thermal effects, whose response is related to the thermal capacity of the interaction volume/waveguide region, has response time-scales many orders longer, or slower, compared to the pump pulse width. 

Since the free charge carrier loss is trivial in the passive semicondutor devices, it would be sufficient for us to model the overall nonlinear loss using a lumped parameter, called two photon absorption coefficient $ \alpha_2 $, which is also wavelength dependent. In the $ \chi^{(2)} $ nonlinearity, the pump wavelength is close to the energy gap of semiconductor material. Hence the pump $ \alpha_2 $ is three orders magnitude higher than that at signal wavelength, and it has the value of $ \alpha_2 = 1.2 \times 10^{-10} $~\SI{}{W}{m}. To determine the value of the nonlinear loss coefficient $ \alpha _{NL} $ defined by
\begin{equation}\label{eq:alpha_L}
{\alpha _{NL}} = {\alpha _2} I_p
\end{equation}
We need to find the pump field intensity $ I_p (0) $, which can be found via the external pump power $ P_p $ and optical coupling efficient $ \eta_{oc} = 65\% $ as
\begin{equation}\label{eq:I_p}
{I_p}(0) = {{{P_{pk}}} \over {{A_{eff}}}}\sqrt {{\epsilon _{eff}}} {\eta _{oc}}
\end{equation}
where $ \epsilon _{eff} = 3.14 $ is the effective refractive index at the pump Bragg mode, whose mode area is $ A_{eff} =6 \times 10^{-12}$\SI{}{m^2}. The peak power of the pulsed pump is computed from the average pump power $ P_{av} $ which was varied up to 15 mW using the relationship
\begin{equation}\label{eq:P_pk}
{P_{pk}} = {{{P_{av}}} \over {f_{rep} \tau }}
\end{equation}
where the pulsed pump repetition rate $ f_{rep} = $ \SI{80}{MHz}. The pulse width $ \tau =$\SI{120}{fs}. 

\begin{equation}\label{eq:I_eq_pz}
\frac{dI_p(z)}{dz}=-\left(\alpha_L+\alpha_2I_p (z)\right)I_p (z)
\end{equation}

\begin{equation}\label{eq:Ip_z}
I_p\left(z\right)=\frac{I_p\left(0\right) \alpha_L e^{-\alpha_Lz}}{\alpha_L+ \alpha_2I_p\left(0\right)\left[1-e^{-\alpha_L z}\right]}	
\end{equation}

Then we use the standard effective length definition \cite{Agrawal2001} to determine the value of $ L_{eff} $ by solving the following nonlinear equation  as:
\begin{equation}\label{eq:L_eff}
{L_{eff}} = {{1 - {e^{ - [{\alpha_L+\alpha_2I_p (L)}]L}}} \over {{\alpha_L+\alpha_2I_p (L_{eff})}}}
\end{equation}
where the physical length of a given device $ L $ is also taken into account is Eq.~(\ref{eq:L_eff}).
\begin{figure}[h]
	\centering
	\includegraphics[draft=false, width=0.4 \columnwidth]{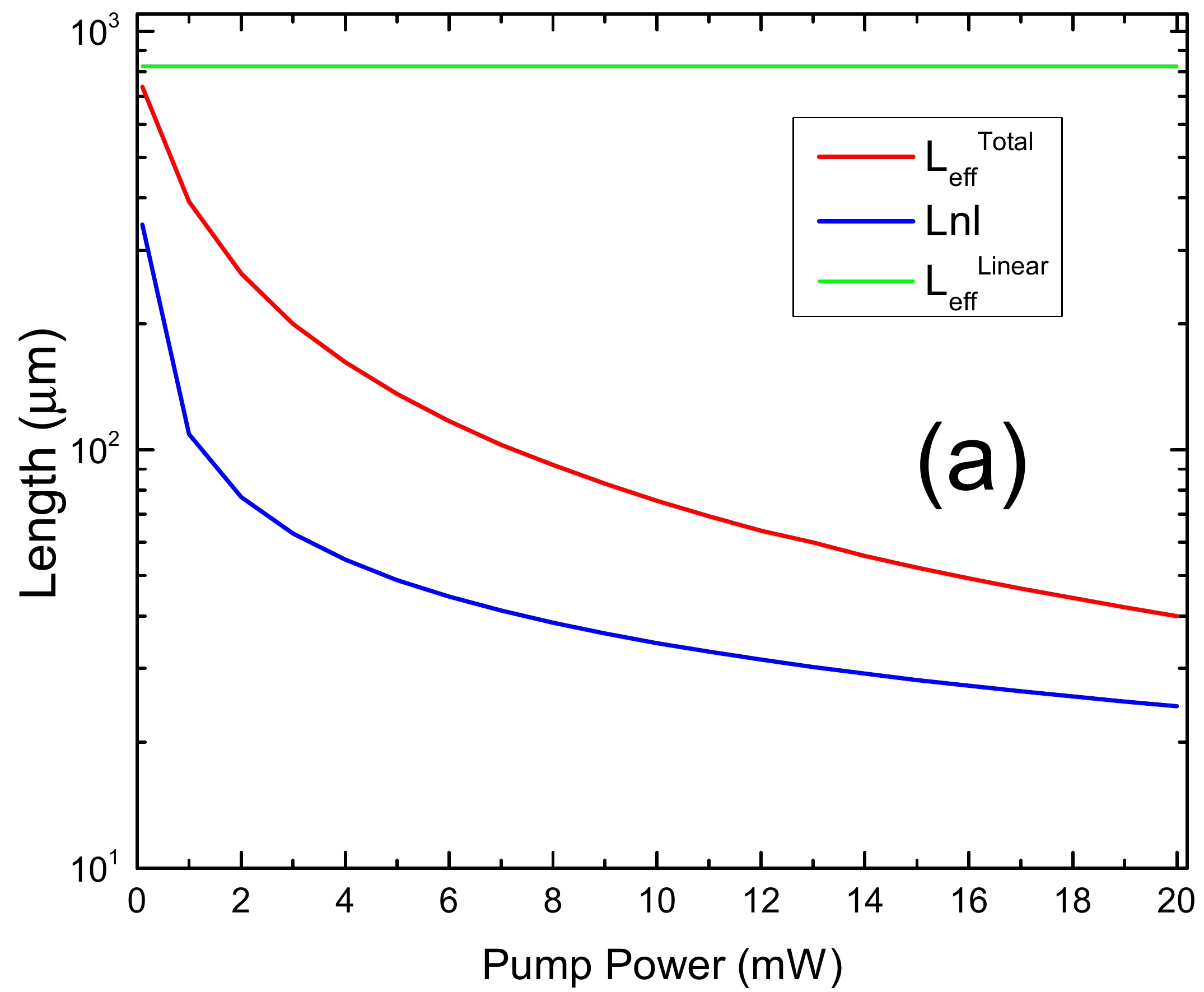}
	\includegraphics[draft=false, width=0.4 \columnwidth]{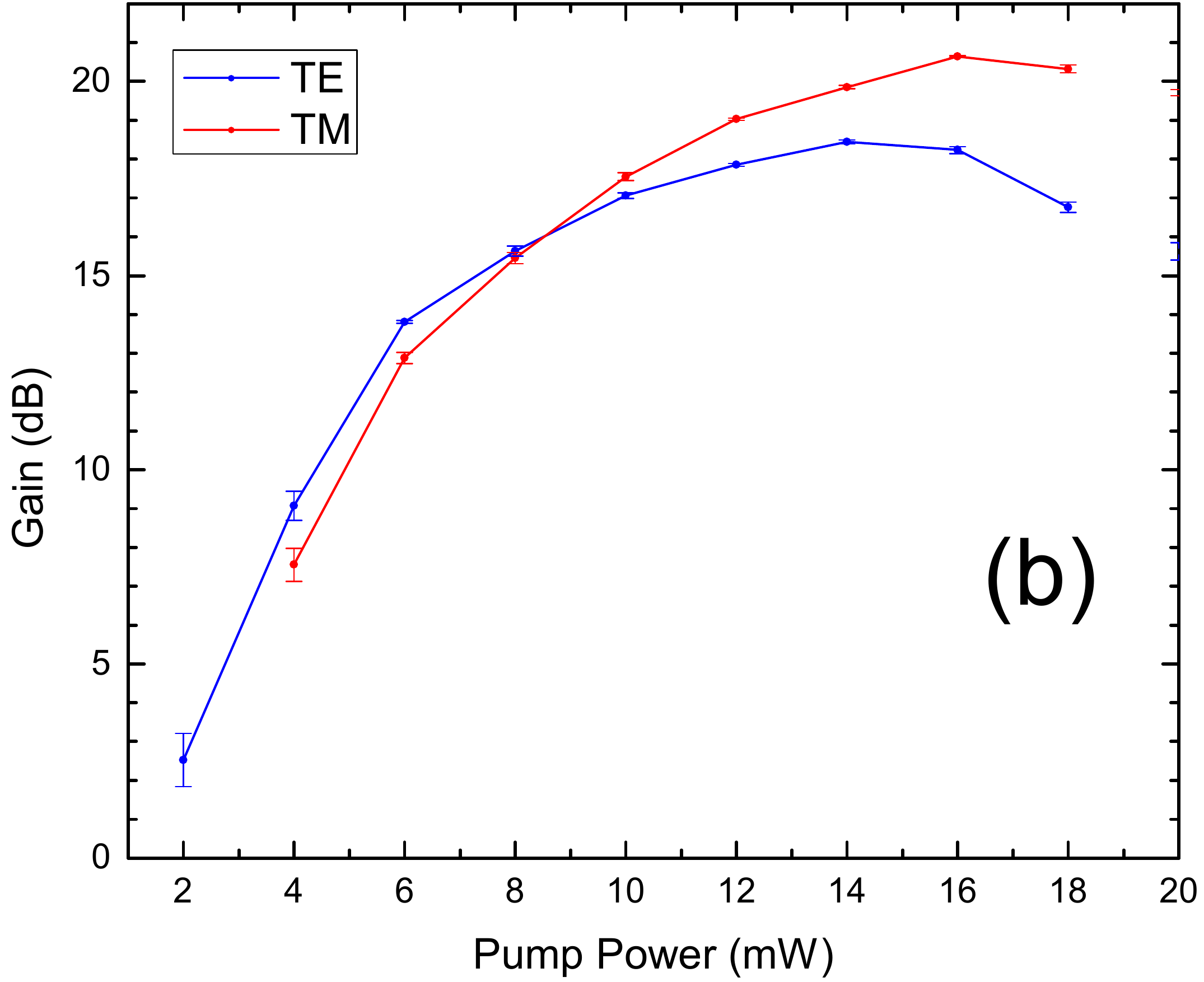}
	\caption{(a) The total effective length (red solid line) dictated by the linear loss (green solid line) $ \alpha_L $ and nonlinear optical loss $ \alpha _t $ in a 1 mm long BRW waveguide pumped by a pulsed 80 MHz repetition rate, and pulse width is 140 femtosecond. The nonlinear length is also plotted (blue solid line) as a function of pump power. The pump wavelength is assume to be 780 nm.  (b) The OPA gain measured by TE and TM input probing signal under various of pump power from 2 mW to 18 mW.
	}
	\label{sfig:Fig-2}
\end{figure}

Based on the above equation, we solved the equation numerically and plotted the effective length $ L_{eff} $ for our devices with physical length $ L=1 $mm in Fig.~\ref{sfig:Fig-2}~(a) as a function of the average pump power $ P_{av} $ from 0 to 15 mW.

The nonlinear length $ L_{NL} $ is associated with the magnitude of $ \chi^{(2)} $ nonlinearity and peak pump power. Most of the parametric gain occurs within this length, which is defined as \cite{Wasilewski2006}: 
\begin{equation}\label{eq:L_NL}
{L_{NL}} = {{8{c^2}{k_s}} \over {\omega _p^2{d_{{\rm{eff}}}}{{\rm{E}}_0}}}
\end{equation}
where $ d_{\rm{eff}} = 200 \rm{pm} V^{-1} $  
is the effective $ \chi^{(2)} $ nonlinearity. 
$ \text{\textbf{E}}_0 $ is the peak pump field.

\begin{equation}\label{eq:E0}
{{\rm{\textbf{E}}}_0} = {\left( {{{{I_p}{\eta _0}} \over {\sqrt {{\varepsilon _{eff}}} }}} \right)^{{1 \over 2}}}
\end{equation}
We evaluated the nonlinear length $ L_{NL} $ within the same pump power range as the TPA. The result is co-plotted in the Fig.~\ref{sfig:Fig-2}~(a). It is evident that the effective length is longer than the nonlinear length.  Since the effective length of the devices is dictated by the nonlinear length, in the experiment, further increasing the pump power above 15 mW will result in no net increase in the gain because the nonlinear loss dominates.  

Fig.~\ref{sfig:Fig-2}~(b) shows the gain measured in both TE and TM input modes as a function of pump power. The curve indicates that both TE and TM parametric gain increases when the pump power increases till 15 mW. After this point, the TE gain starts to roll over downward faster than that in the TM mode, but both modes exhibited the same trend, because the nonlinear loss is more severe in the TE mode than that in the TM mode, leading to the effective length degrades faster in the TE mode.

The short effective length allows us to implement such amplifiers in highly compact form factor for highly integrated photonic circuits. The integration within shorter length also leads to the lowest photon losses possible. 

\ans{The propagation loss for the short wavelength, pump, is estimated to be 17 dB/cm. The  propagation loss for the signal is about 2 dB/cm. The loss at the pump poses a limit on the effective length on the order of 0.85 mm. While this is a short length for these devices, the effective length dictated by the nonlinear loss is on the order of 0.050 mm to 0.2 mm scale. Therefore the limit on the nonlinear interaction in our devices is posed by the nonlinear losses despite of the high linear losses. }

\setcounter{equation}{0}
\setcounter{figure}{0}
\section{multimode squeezing formalism to model optical parametric amplification}
Quantum mechanically, the single spatial mode type-II OPA process can be modeled as multiple channels (labeled by \(k\)) of two-mode squeezing of TE and TM polarized time-frequency modes. Each channel is associated with a squeezing parameter \(r_k\) that dictates the corresponding OPA gain. The Bogoliubov transform of the OPA input/output mode operators are given by:
\begin{gather}
\hat A_{TE}^{(k,\text{out})} = \cosh ({r_k})\hat A_{TE}^{(k,\text{in})}  + \sinh ({r_k})\hat A_{TM}^{(k,\text{in})} \\
\hat A_{TM}^{(k,\text{out})} = \cosh ({r_k})\hat A_{TM}^{(k,\text{in})}  + \sinh ({r_k})\hat A_{TE}^{(k,\text{in})}
\end{gather}
where \(\hat A_{p}^{(k,\text{io})}, p=TE,TM, io = in,out\) is the \(k\)th input (output), TE (TM) mode operator. To achieve largest possible amplification, the input signal light of either TE or TM polarization should be in the mode with the largest squeezing parameter \(r_0\) (designate \(k=0\) for this mode). However, the time frequency amplitude profile associated with such mode is in general complicated depending on the OPA pump light spectrum as well as the waveguide structure. As a result, it is experimentally challenging to selectively couple light into the largest squeezing mode. In this work, we take an alternative approach by coupling the signal light equally into all TE or TM time frequency modes (i.e. continuous wave signal light and ultrafast pump light) and then estimate the amount of power \(P_{0,TE}\)(\(P_{0,TM}\)) coupled into the largest squeezing mode. Then the OPA gain corresponding to the largest squeezing mode can be determined by taking the ratio between the \(P_{0,TE}\)(\(P_{0,TM}\)) and the output power.

\setcounter{equation}{0}
\setcounter{figure}{0}
\section{Experimental steps for determining the OPA input power}
To accurately determine the input power that is coupled into the strongest squeezing mode of the OPA, we perform a three-stage experiment. First, we determine through modeling and experimental characterization that the semiconductor waveguide device under test provides amplification predominantly on a single pair (TE and TM polarization) of  time-frequency modes. Second, we characterize the OPA gain factor under a low pump power (2mW) condition through photon number correlation measurement. The last step is to measure the amplified signal power with 2mW pump. Then the input signal power coupled into the maximally amplifying mode can then be determined as the ratio between the amplified signal power and the OPA gain under the low pump condition.

\subsection{The cross-correlation function $g^{(1,1)}$ and its relationship to squeezing}
When an OPA is pumped without input signal, spontaneous parametric down conversion (SPDC) of photon pairs will occur. The photon counting and correlation statistics of the generated photon pairs allows indirect determination of the OPA gain without any input light. In particular, for a type-II OPA, the cross correlation \(g^{(1,1)}\) between the output TE and TM mode and the autocorrelation \(g^{(2)}\) of either the output TE or TM mode, is related to the squeezing parameters of all two-mode squeezing channels as \cite{Christ2011}:  
\begin{equation}\label{eq:g11}
{g^{\left( {1,1} \right)}} = {g^{\left( 2 \right)}} + \frac{1}{{\sum\nolimits_k {{{\sinh }^2} {{r_k}} } }}
\end{equation}
The first order cross correlation \(g^{(1,1)}\) can be experimentally determined as the ratio between the coincidence detection rate and single channel detection rate of TE and TM output photons. The auto-correlation \(g^{(2)}\) can be measured by splitting either the TE or TM mode into two independent photo detectors and perform coincidence detection in principle. Since the  \(g^{(2)}\)  is not loss resilient, it will likely be altered due to the high optical loss (4~\% of overall optical efficiency) of our measurement setup. As a result, \(g^{(2)}\) value can not be the actual value dictated by our theory.

Experiment result shows that with 2mW power power,\(g^{(1,1)}=6.8 \), comparing with the theoretical value of \(g^{(1,1)}=7.4 \); In addition,  we also theoretically calculated \(g^{(2)}= 1.2\) due to the reason aforementioned. The corresponding \(\sum\nolimits_k {{{\sinh }^2} {{r_k}} }=0.165\). For simplicity of notation, we define the lumped squeezing parameter \(r_{tot} = \arcsinh(\sqrt{\sum_k\sinh^2r_k})\).

\begin{figure*}[!h]
	\normalsize
	\begin{equation}\label{eq:delta_k}
	\Delta k \simeq \left(\beta_{ 1p}-\beta_{1}\right)\left(\omega_s+\omega_i-\omega_{p}\right) -\frac{1}{2} \beta_{2}\left[\left(\omega_s-\omega_{p} / 2\right)^{2}   
	+\left(\omega_i-\omega_{p} / 2\right)^{2}\right] +\frac{1}{2} \beta_{2 p}\left(\omega_s+\omega_i-\omega_{p}\right)^{2} 
	\end{equation}
	\hrulefill
\end{figure*}

\subsection{The Single Schmidt mode approximation}
We confirm whether our OPA is operated in a nearly single mode by calculating the number of Schmidt modes (Schmidt number) that corresponds to the JSA. The computation of JSA is comprised of the expression of the expected phase matching profile $
\operatorname{sinc} \frac{ \Delta k  L_{eff}}{2} $ is dictated by two factors, i.e. the $ \Delta k=k_{p}\left(\omega_s+\omega_i\right)-k_s(\omega_s)-k_i\left(\omega_i \right) $ and the device effective length $ L_{eff} $. 

We can determine the $ \Delta k $ via numerical approximation as \cite{Wasilewski2006} in Eq.~(C. 2), where $\beta_{n}=\left.\frac{d^{n} k(\omega)}{d \omega^{n}}\right|_{\omega=\omega_{p} / 2}$; and $\beta_{np}=\left.\frac{d^{n} k_p(\omega)}{d \omega^{n}}\right|_{\omega=\omega_{p}}$. For the device effective length $ L_{eff} $, please refer to the section I for details.

\begin{figure}
	\centering
	\includegraphics[draft=false, width=0.45 \columnwidth]{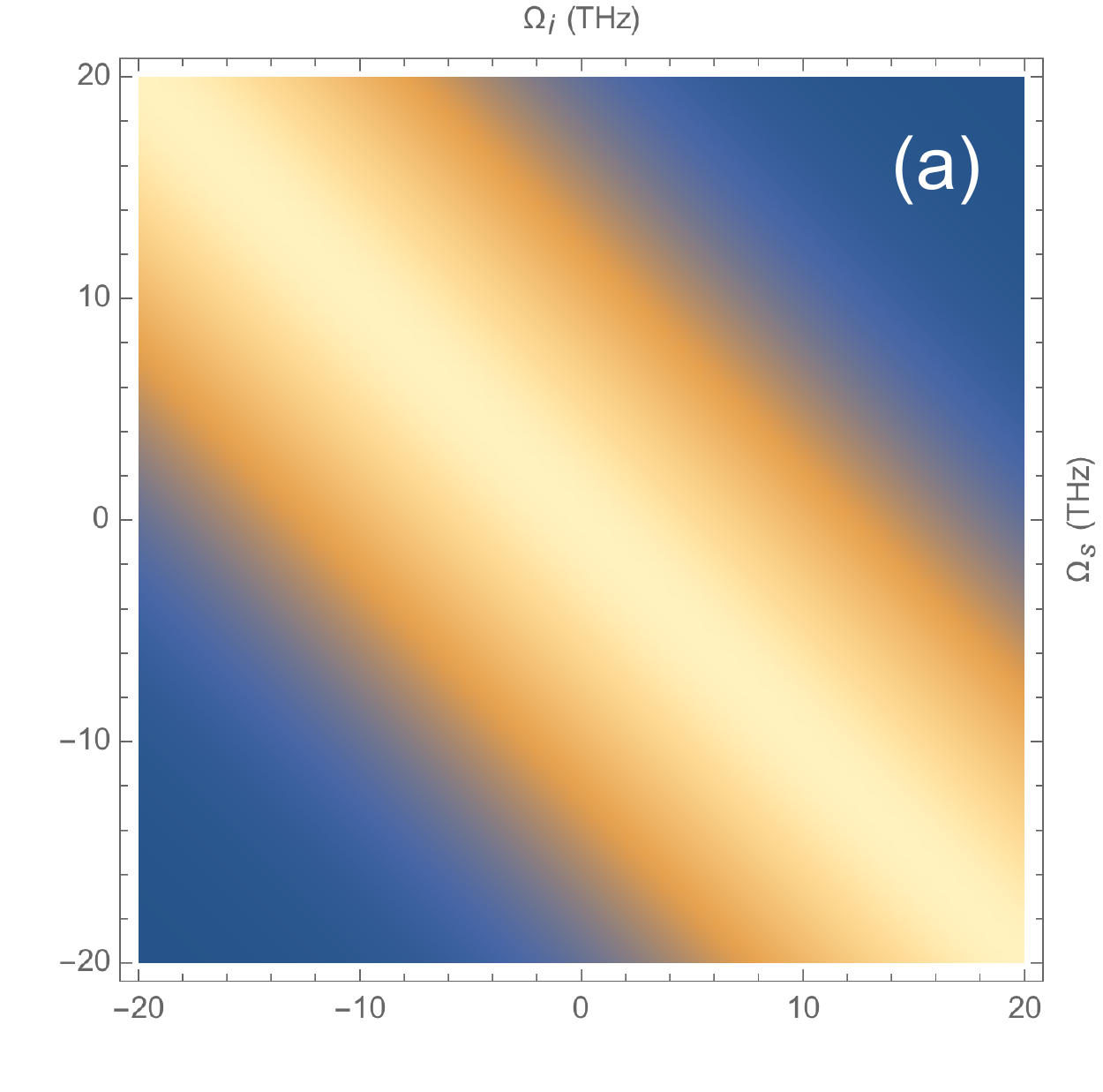}
	\includegraphics[draft=false, width=0.5 \columnwidth]{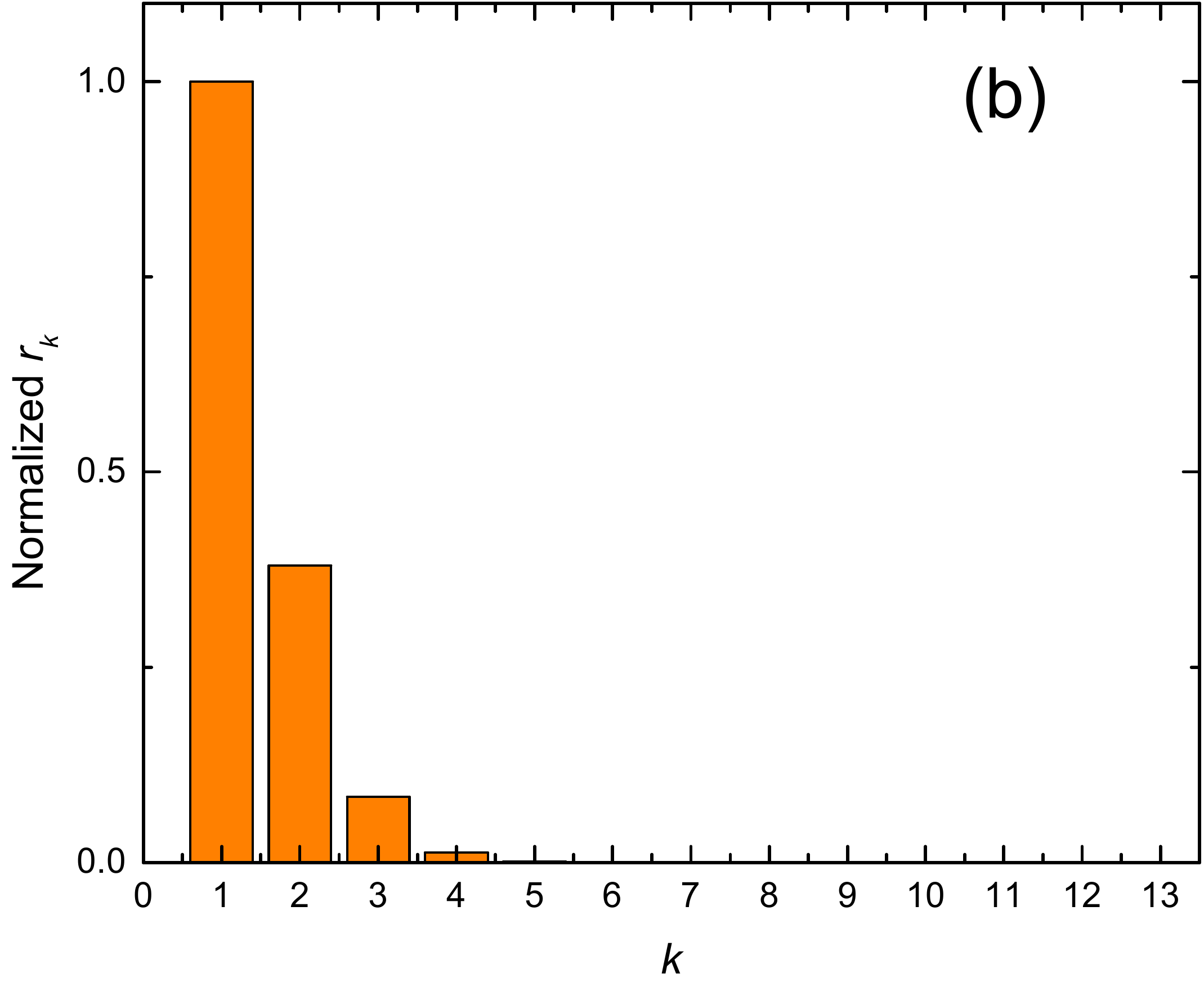}
	\caption{(a)Joint spectrum amplitude (JSA) of our BRW waveguide pumped at 15 mW average power. (b) Schmidt decomposition displaying the normalized $ r_k $ of parallel Schmidt modes.
	}
	\label{sfig:Schmidt}
\end{figure}

Note that JSA that is used here corresponds to the state that has passed through an equivalent spectral filter by about 200 nm bandwidth by virtue of the setup and detection spectral response. The JSA is centered at 1558 nm. The JSA is plotted in Fig.~\ref{sfig:Schmidt}~(a).

By Schmidt decomposition method, we are able to determine the gain of all parallel OPAs using Bloch-Messiah reduction theory and the results is plotted in Fig.~\ref{sfig:Schmidt}~(b). 

The Schmidt number $ K $ is 1.34. The entropy $ S=0.65 $ is close to zero. We can interpret this entropy as approximately close to, but not exactly, a single Schmidt mode. For us to quantify how close it is, we found this value $ r_1/r_{\text{tot}} = 93~\% $, which is the ratio of the first parallel OPA gain $ r_1 $ over the total squeezing parameter $ r_{\text{tot}} $.

Note, that another approximate indication of the temporal mode number can be estimated through the product of pump pulse width  $ \tau_\text{pulse} $=100 fs  and phase matching bandwidth $ B $=1.2 THz, which is $ B \tau_\text{pulse}=0.12 $ for our measurements. This further suggests that the number of temporal mode being amplified are approximately one.

\subsection{Determining input power $P_S(0) $ Via $r_{\text{tot}}$}
Having determined \(r_{tot}=0.4\) for 2mW pump power, it is possible to determine the power \(P_{0,TE}\)(\(P_{0,TM}\)) of input signal light that is coupled into the amplified mode by measuring the amplified output. Assuming the input signal is either TE or TM and drop the subscript \(p\).  Then the total output power from the OPA could be expressed as:   
\begin{gather}
P_{out} = \sum_k(P_k+ P_k\sinh^2(r_k))+ P_{sp}\\
= \sum_k(P_k+ P_k\sinh^2(r_k))+ P_{sp}
\end{gather}
where $ P_{sp} $ refers to the spontaneous emission power and all background lights that is coupled to the collection optics.

Since the OPA is pumped with ultrafast pump light and the signal light is continuous wave, it is safe to assume that the input power \(P_k\) in each two-mode squeezing channel is identical (\(P_k=P_0\)). Then the total output signal power \(P_s(L)\) is given by:
\begin{gather}
P_s(L) = \sum_k(P_k+ P_k\sinh^2(r_k))+P_{sp}\\
= \sum_kP_k+ P_0\sinh^2(r_{tot}))+ P_{sp}	\label{eq.appendix.PsL}
\end{gather}
In the experiment, the total amplified power \(P_{out}\) is measured through lock-in amplification with the pump light being periodically modulated by a mechanical chopper. Therefore, the first term of the above expression that is due to the linear transmission of the input power is not detected. The last term $ P_{sp} $ due to the OPA SPDC power can be determined by turning off the signal input (with only pump light on). 

After excluding these two terms, the pump power increases while varying the signal wavelength to characterize the wavelength dependent gain. Note that in order to make $ P_s(L) = P_0 \cosh^2(r_{tot})  $ valid, we have over 93~\% fidelity that we can approximate our OPA to be in single mode. 

We can determine the input power of the amplifying mode are given by \(P_{0}^{TE}=3.1\times10^{-11}\)~W for TE input signal and  \(P_{0}^{TM}=3.0\times10^{-11}\)~W for TM input signal. After determining the value of \(P_{0}^{TE}\) and \(P_{0}^{TM}\), the pump is gradually increased up to 20~mW to characterize the OPA gain. However, we found the optimal pump is at 15~mW.

\bibliographystyle{IEEEtran}

\begin{thebibliography}{10}
	\providecommand{\url}[1]{#1}
	\csname url@samestyle\endcsname
	\providecommand{\newblock}{\relax}
	\providecommand{\bibinfo}[2]{#2}
	\providecommand{\BIBentrySTDinterwordspacing}{\spaceskip=0pt\relax}
	\providecommand{\BIBentryALTinterwordstretchfactor}{4}
	\providecommand{\BIBentryALTinterwordspacing}{\spaceskip=\fontdimen2\font plus
		\BIBentryALTinterwordstretchfactor\fontdimen3\font minus
		\fontdimen4\font\relax}
	\providecommand{\BIBforeignlanguage}[2]{{%
			\expandafter\ifx\csname l@#1\endcsname\relax
			\typeout{** WARNING: IEEEtran.bst: No hyphenation pattern has been}%
			\typeout{** loaded for the language `#1'. Using the pattern for}%
			\typeout{** the default language instead.}%
			\else
			\language=\csname l@#1\endcsname
			\fi
			#2}}
	\providecommand{\BIBdecl}{\relax}
	\BIBdecl
	
	\bibitem{Wang1965}
	\BIBentryALTinterwordspacing
	C.~C. Wang and G.~W. Racette, ``{Measurement of parametric gain accompanying
		optical difference frequency generation},'' \emph{Applied Physics Letters},
	vol.~6, no.~8, pp. 169--171, apr 1965. [Online]. Available:
	\url{https://doi.org/10.1063/1.1754219
		http://aip.scitation.org/doi/10.1063/1.1754219}
	\BIBentrySTDinterwordspacing
	
	\bibitem{Levenson1993}
	J.~A. Levenson, I.~Abram, T.~Rivera, and P.~Grangier, ``{Reduction of quantum
		noise in optical parametric amplification},'' \emph{Journal of the Optical
		Society of America B}, vol.~10, no.~11, p. 2233, 1993.
	
	\bibitem{Hansryd2002}
	J.~Hansryd, P.~A. Andrekson, M.~Westlund, J.~Li, and P.~O. Hedekvist,
	``{Fiber-based optical parametric amplifiers and their applications},''
	\emph{IEEE Journal on Selected Topics in Quantum Electronics}, vol.~8, no.~3,
	pp. 506--520, 2002.
	
	\bibitem{Foster2006}
	\BIBentryALTinterwordspacing
	M.~A. Foster, A.~C. Turner, J.~E. Sharping, B.~S. Schmidt, M.~Lipson, and A.~L.
	Gaeta, ``{Broad-band optical parametric gain on a silicon photonic chip},''
	\emph{Nature}, vol. 441, no. 7096, pp. 960--963, jun 2006. [Online].
	Available: \url{http://dx.doi.org/10.1038/nature04932
		http://www.nature.com/doifinder/10.1038/nature04932}
	\BIBentrySTDinterwordspacing
	
	\bibitem{Salem2008}
	R.~Salem, M.~a. Foster, A.~C. Turner, D.~F. Geraghty, M.~Lipson, and A.~L.
	Gaeta, ``{Signal regeneration using low-power four-wave mixing on silicon
		chip},'' \emph{Nature Photonics}, vol.~2, pp. 35--38, 2008.
	
	\bibitem{Liu2010}
	X.~Liu, R.~M. Osgood, Y.~a. Vlasov, and W.~M.~J. Green, ``{Mid-infrared optical
		parametric amplifier using silicon nanophotonic waveguides},'' \emph{Nature
		Photonics}, vol.~4, no.~8, pp. 557--560, 2010.
	
	\bibitem{Kuyken2011}
	B.~Kuyken, X.~Liu, G.~Roelkens, R.~Baets, R.~M. {Osgood, Jr.}, and W.~M.~J.
	Green, ``{50 dB parametric on-chip gain in silicon photonic wires},''
	\emph{Optics Letters}, vol.~36, no.~22, pp. 4401--4403, 2011.
	
	\bibitem{Liu2012}
	\BIBentryALTinterwordspacing
	X.~Liu, B.~Kuyken, G.~Roelkens, R.~Baets, R.~M. Osgood, and W.~M.~J. Green,
	``{Bridging the mid-infrared-to-telecom gap with silicon nanophotonic
		spectral translation},'' \emph{Nature Photonics}, vol.~6, no.~10, pp.
	667--671, 2012. [Online]. Available:
	\url{http://dx.doi.org/10.1038/nphoton.2012.221}
	\BIBentrySTDinterwordspacing
	
	\bibitem{Schroder2013}
	\BIBentryALTinterwordspacing
	J.~Schroder, R.~Neo, Y.~Paquot, D.-Y. Choi, S.~Madden, B.~Luther-Davies, and
	B.~J. Eggleton, ``{Phase-sensitive amplification in a $\chi$(3) photonic
		chip},'' in \emph{2013 Conference on Lasers {\&} Electro-Optics Europe {\&}
		International Quantum Electronics Conference CLEO EUROPE/IQEC}.\hskip 1em
	plus 0.5em minus 0.4em\relax IEEE, may 2013, pp. 1--1. [Online]. Available:
	\url{http://ieeexplore.ieee.org/document/6800824/}
	\BIBentrySTDinterwordspacing
	
	\bibitem{Lamont2008}
	M.~R.~E. Lamont, B.~Luther-Davies, D.-Y. Choi, S.~Madden, X.~Gai, and B.~J.
	Eggleton, ``{Net-gain from a parametric amplifier on a chalcogenide optical
		chip},'' \emph{Optics Express}, vol.~16, no.~25, pp. 20\,374--20\,381, 2008.
	
	\bibitem{Kim1992}
	\BIBentryALTinterwordspacing
	D.-S. Kim, J.~Shah, T.~C. Damen, W.~Sch{\"{a}}fer, F.~Jahnke, S.~Schmitt-Rink,
	and K.~K{\"{o}}hler, ``{Unusually slow temporal evolution of femtosecond
		four-wave-mixing signals in intrinsic GaAs quantum wells: Direct evidence for
		the dominance of interaction effects},'' \emph{Physical Review Letters},
	vol.~69, no.~18, pp. 2725--2728, nov 1992. [Online]. Available:
	\url{http://link.aps.org/doi/10.1103/PhysRevLett.69.2013
		http://link.aps.org/doi/10.1103/PhysRevLett.70.694
		http://link.aps.org/doi/10.1103/PhysRevLett.69.2725}
	\BIBentrySTDinterwordspacing
	
	\bibitem{Dolgaleva2015}
	\BIBentryALTinterwordspacing
	K.~Dolgaleva, P.~Sarrafi, P.~Kultavewuti, K.~M. Awan, N.~Feher, J.~S.
	Aitchison, L.~Qian, M.~Volatier, R.~Ar{\`{e}}s, and V.~Aimez, ``{Tuneable
		four-wave mixing in AlGaAs nanowires},'' \emph{Optics Express}, vol.~23,
	no.~17, pp. 22\,477--22\,493, 2015. [Online]. Available:
	\url{http://www.osapublishing.org/viewmedia.cfm?uri=oe-23-17-22477{\&}seq=0{\&}html=true}
	\BIBentrySTDinterwordspacing
	
	\bibitem{Wathen2014}
	\BIBentryALTinterwordspacing
	J.~J. Wathen, P.~Apiratikul, C.~J.~K. Richardson, G.~a. Porkolab, G.~M. Carter,
	and T.~E. Murphy, ``{Efficient continuous-wave four-wave mixing in
		bandgap-engineered AlGaAs waveguides.}'' \emph{Optics letters}, vol.~39,
	no.~11, pp. 3161--4, 2014. [Online]. Available:
	\url{http://ol.osa.org/abstract.cfm?URI=ol-39-11-3161}
	\BIBentrySTDinterwordspacing
	
	\bibitem{Chang2020}
	\BIBentryALTinterwordspacing
	L.~Chang, W.~Xie, H.~Shu, Q.-F. Yang, B.~Shen, A.~Boes, J.~D. Peters, W.~Jin,
	C.~Xiang, S.~Liu, G.~Moille, S.-P. Yu, X.~Wang, K.~Srinivasan, S.~B. Papp,
	K.~Vahala, and J.~E. Bowers, ``{Ultra-efficient frequency comb generation in
		AlGaAs-on-insulator microresonators},'' \emph{Nature Communications},
	vol.~11, no.~1, p. 1331, dec 2020. [Online]. Available:
	\url{https://doi.org/10.1038/s41467-020-15005-5
		http://www.nature.com/articles/s41467-020-15005-5}
	\BIBentrySTDinterwordspacing
	
	\bibitem{Ooi2017}
	\BIBentryALTinterwordspacing
	K.~J.~A. Ooi, D.~K.~T. Ng, T.~Wang, A.~K.~L. Chee, S.~K. Ng, Q.~Wang, L.~K.
	Ang, A.~M. Agarwal, L.~C. Kimerling, and D.~T.~H. Tan, ``{Pushing the limits
		of CMOS optical parametric amplifiers with USRN:Si7N3 above the two-photon
		absorption edge},'' \emph{Nature Communications}, vol.~8, p. 13878, jan 2017.
	[Online]. Available:
	\url{http://www.nature.com/doifinder/10.1038/ncomms13878}
	\BIBentrySTDinterwordspacing
	
	\bibitem{Umeki2011}
	T.~Umeki, O.~Tadanaga, A.~Takada, and M.~Asobe, ``{Phase sensitive degenerate
		parametric amplification using directly-bonded PPLN ridge waveguides.}''
	\emph{Optics express}, vol.~19, no.~7, pp. 6326--6332, 2011.
	
	\bibitem{Umeki2013a}
	\BIBentryALTinterwordspacing
	T.~Umeki, M.~Asobe, and H.~Takenouchi, ``{In-line phase sensitive amplifier
		based on PPLN waveguides.}'' \emph{Optics express}, vol.~21, no.~10, pp.
	12\,077--84, 2013. [Online]. Available:
	\url{http://www.opticsexpress.org/abstract.cfm?URI=oe-21-10-12077}
	\BIBentrySTDinterwordspacing
	
	\bibitem{Ishimoto2016}
	T.~A.~K. Ishimoto, K.~O. J. I.~I. Nafune, Y.~O. H.~O. Gawa, and H.~I.~S. Asaki,
	``{Highly efficient phase-sensitive parametric gain in periodically poled
		LiNbO 3 ridge waveguide},'' \emph{Optics Letters}, vol.~41, no.~9, pp.
	1905--1908, 2016.
	
	\bibitem{Ledezma:22}
	\BIBentryALTinterwordspacing
	L.~Ledezma, R.~Sekine, Q.~Guo, R.~Nehra, S.~Jahani, and A.~Marandi, ``{Intense
		optical parametric amplification in dispersion-engineered nanophotonic
		lithium niobate waveguides},'' \emph{Optica}, vol.~9, no.~3, p. 303, mar
	2022. [Online]. Available:
	\url{http://opg.optica.org/optica/abstract.cfm?URI=optica-9-3-303
		https://opg.optica.org/abstract.cfm?URI=optica-9-3-303}
	\BIBentrySTDinterwordspacing
	
	\bibitem{Jankowski:22}
	\BIBentryALTinterwordspacing
	M.~Jankowski, N.~Jornod, C.~Langrock, B.~Desiatov, A.~Marandi, M.~Lon{\v{c}}ar,
	and M.~M. Fejer, ``{Quasi-static optical parametric amplification},''
	\emph{Optica}, vol.~9, no.~3, p. 273, mar 2022. [Online]. Available:
	\url{http://opg.optica.org/optica/abstract.cfm?URI=optica-9-3-273
		https://opg.optica.org/abstract.cfm?URI=optica-9-3-273}
	\BIBentrySTDinterwordspacing
	
	\bibitem{Ravaro2007}
	M.~Ravaro, M.~{Le D{\^{u}}}, J.~P. Likforman, S.~Ducci, V.~Berger, G.~Leo, and
	X.~Marcadet, ``{Estimation of parametric gain in GaAs/AlOx waveguides by
		fluorescence and second harmonic generation measurements},'' \emph{Applied
		Physics Letters}, vol.~91, no.~19, pp. 10--13, 2007.
	
	\bibitem{Ozanam2014}
	C.~Ozanam, M.~Savanier, L.~Lanco, X.~Lafosse, G.~Almuneau, A.~Andronico,
	I.~Favero, S.~Ducci, and G.~Leo, ``{Toward an AlGaAs/AlOx near-infrared
		integrated optical parametric oscillator},'' \emph{Journal of the Optical
		Society of America B-optical Physics}, vol.~31, no.~3, pp. 542--550, 2014.
	
	\bibitem{Abolghasem2012}
	P.~Abolghasem, J.~B. Han, D.~Kang, B.~J. Bijlani, and A.~S. Helmy,
	``{Monolithic photonics using second-order optical nonlinearities in
		multilayer-core bragg reflection waveguides},'' \emph{IEEE Journal on
		Selected Topics in Quantum Electronics}, vol.~18, no.~2, pp. 812--825, 2012.
	
	\bibitem{Han2009}
	\BIBentryALTinterwordspacing
	J.~Han, P.~Abolghasem, B.~J. Bijlani, and A.~S. Helmy, ``{Continuous-wave
		sum-frequency generation in AlGaAs Bragg reflection waveguides},''
	\emph{Optics Letters}, vol.~34, no.~23, p. 3656, dec 2009. [Online].
	Available: \url{https://www.osapublishing.org/abstract.cfm?URI=ol-34-23-3656}
	\BIBentrySTDinterwordspacing
	
	\bibitem{Han2010a}
	\BIBentryALTinterwordspacing
	J.~B. Han, D.~P. Kang, P.~Abolghasem, B.~J. Bijlani, and A.~S. Helmy,
	``{Pulsed- and continuous-wave difference-frequency generation in AlGaAs
		Bragg reflection waveguides},'' \emph{Journal of the Optical Society of
		America B}, vol.~27, no.~12, p. 2488, dec 2010. [Online]. Available:
	\url{https://www.osapublishing.org/abstract.cfm?URI=ol-34-23-3656
		https://www.osapublishing.org/abstract.cfm?URI=josab-27-12-2488}
	\BIBentrySTDinterwordspacing
	
	\bibitem{Abolghasem2009}
	\BIBentryALTinterwordspacing
	P.~Abolghasem, {Junbo Han}, B.~Bijlani, A.~Arjmand, and A.~Helmy, ``{Highly
		Efficient Second-Harmonic Generation in Monolithic Matching Layer Enhanced
		Al{\$}{\_}x{\$}Ga{\$}{\_}{\{}1-x{\}}{\$}As Bragg Reflection Waveguides},''
	\emph{IEEE Photonics Technology Letters}, vol.~21, no.~19, pp. 1462--1464,
	oct 2009. [Online]. Available:
	\url{http://ieeexplore.ieee.org/document/5173550/}
	\BIBentrySTDinterwordspacing
	
	\bibitem{Caves1981}
	\BIBentryALTinterwordspacing
	C.~M. Caves, ``{Quantum limits on noise in linear amplifiers},'' \emph{Physical
		Review D}, vol.~26, no.~8, pp. 1817--1839, oct 1982. [Online]. Available:
	\url{https://link.aps.org/doi/10.1103/PhysRevD.26.1817}
	\BIBentrySTDinterwordspacing
	
	\bibitem{Han2010}
	\BIBentryALTinterwordspacing
	J.~B. Han, P.~Abolghasem, B.~J. Bijlani, A.~Arjmand, S.~C. Kumar,
	A.~Esteban-Martin, M.~Ebrahim-Zadeh, and A.~S. Helmy, ``{Femtosecond
		second-harmonic generation in AlGaAs Bragg reflection waveguides: theory and
		experiment},'' \emph{Journal of the Optical Society of America B}, vol.~27,
	no.~6, p. 1291, jun 2010. [Online]. Available:
	\url{https://www.osapublishing.org/abstract.cfm?URI=josab-27-6-1291}
	\BIBentrySTDinterwordspacing
	
	\bibitem{Yan:22}
	\BIBentryALTinterwordspacing
	Z.~Yan, H.~He, H.~Liu, M.~Iu, O.~Ahmed, E.~Chen, P.~Blakey, Y.~Akasaka,
	T.~Ikeuchi, and A.~S. Helmy, ``A $\chi^{(2)} $ -based \text{AlGaAs} phase
	sensitive amplifier with record gain, noise, and sensitivity,''
	\emph{Optica}, vol.~9, no.~1, p.~56, jan 2022. [Online]. Available:
	\url{http://opg.optica.org/optica/abstract.cfm?URI=optica-9-1-56
		https://opg.optica.org/abstract.cfm?URI=optica-9-1-56}
	\BIBentrySTDinterwordspacing
	
	\bibitem{Levenson1997}
	\BIBentryALTinterwordspacing
	J.~A. Levenson, K.~Bencheikh, D.~J. Lovering, P.~Vidakovic, and C.~Simonneau,
	``{Quantum noise in optical parametric amplification: a means to achieve
		noiseless optical functions},'' \emph{Quantum and Semiclassical Optics:
		Journal of the European Optical Society Part B}, vol.~9, no.~2, pp. 221--237,
	apr 1997. [Online]. Available:
	\url{http://stacks.iop.org/1355-5111/9/i=2/a=009?key=crossref.38ff829f26658fd98d741aa18e30ca1e}
	\BIBentrySTDinterwordspacing
	
	\bibitem{Richards1994}
	\BIBentryALTinterwordspacing
	P.~L. Richards, ``{Bolometers for infrared and millimeter waves},''
	\emph{Journal of Applied Physics}, vol.~76, no.~1, pp. 1--24, jul 1994.
	[Online]. Available: \url{http://aip.scitation.org/doi/10.1063/1.357128}
	\BIBentrySTDinterwordspacing
	
	\bibitem{Agarwal2014}
	A.~Agarwal, J.~M. Dailey, P.~Toliver, and N.~A. Peters, ``{Entangled-pair
		transmission improvement using distributed phase-sensitive amplification},''
	\emph{Physical Review X}, vol.~4, no.~4, pp. 1--7, 2014.
	
	\bibitem{Aitchison1997}
	\BIBentryALTinterwordspacing
	J.~Aitchison, D.~Hutchings, J.~Kang, G.~Stegeman, and A.~Villeneuve, ``{The
		nonlinear optical properties of AlGaAs at the half band gap},'' \emph{IEEE
		Journal of Quantum Electronics}, vol.~33, no.~3, pp. 341--348, mar 1997.
	[Online]. Available: \url{http://ieeexplore.ieee.org/document/556002/}
	\BIBentrySTDinterwordspacing
	
	\bibitem{Agrawal2001}
	G.~Agrawal, \emph{{Nonlinear Fiber Optics}}, 2001.
	
	\bibitem{Wasilewski2006}
	W.~Wasilewski, A.~I. Lvovsky, K.~Banaszek, and C.~Radzewicz, ``{Pulsed squeezed
		light: Simultaneous squeezing of multiple modes},'' \emph{Physical Review A -
		Atomic, Molecular, and Optical Physics}, vol.~73, no.~6, pp. 1--12, 2006.
	
	\bibitem{Christ2011}
	\BIBentryALTinterwordspacing
	A.~Christ, K.~Laiho, A.~Eckstein, K.~N. Cassemiro, and C.~Silberhorn,
	``{Probing multimode squeezing with correlation functions},'' \emph{New
		Journal of Physics}, vol.~13, no.~3, p. 033027, mar 2011. [Online].
	Available:
	\url{http://stacks.iop.org/1367-2630/13/i=3/a=033027?key=crossref.765ddf84544144ad1304b2ecf9fffa36}
	\BIBentrySTDinterwordspacing
	
\end{thebibliography}

\end{document}